\documentclass{vgtc}                          

\ifpdf
  \pdfoutput=1\relax                   
  \usepackage{graphicx}                
  \DeclareGraphicsExtensions{.pdf,.png,.jpg,.jpeg} 
\else
  \usepackage{graphicx}                
  \DeclareGraphicsExtensions{.eps}     
\fi%

\graphicspath{{figures/}{pictures/}{images/}{./}} 

\usepackage{microtype}                 
\PassOptionsToPackage{warn}{textcomp}  
\usepackage{textcomp}                  
\usepackage{mathptmx}                  
\usepackage{times}                     
\newcommand{\ie}{\textit{i.e.},\xspace}
\newcommand{\etal}{et al.\xspace}

\usepackage{cite}                      
\usepackage{tabu}                      
\usepackage{booktabs}                  
\usepackage{amsmath,amsthm,amssymb}
\usepackage{algorithm}
\usepackage{algpseudocode}
\usepackage{xspace}
\usepackage{multirow}

\onlineid{1008}

\vgtccategory{Research}

\vgtcinsertpkg
\newcommand{\myparagraph}[1]{\vspace{1mm} \noindent \textbf{#1}}

\title{Jacobi Set Driven Search for Flexible Fiber Surface Extraction}

\author{Mohit Sharma\thanks{e-mail: mohitsharma@iisc.ac.in} %
\and Vijay Natarajan\thanks{e-mail: vijayn@iisc.ac.in}} 
\affiliation{Indian Institute of Science, Bangalore}

\abstract{Isosurfaces are an important tool for analysis and visualization of univariate scalar fields. Earlier works have demonstrated the presence of interesting isosurfaces at isovalues close to critical values. This motivated the development of efficient methods for computing individual components of isosurfaces restricted to a region of interest. Generalization of isosurfaces to fiber surfaces and critical points to Jacobi sets has resulted in new approaches for analyzing bivariate scalar fields. Unlike isosurfaces, there exists no output sensitive method for computing fiber surfaces. Existing methods traverse through all the tetrahedra in the domain. In this paper, we propose the use of the Jacobi set to identify fiber surface components of interest and present an output sensitive approach for its computation. The Jacobi edges are used to initiate the search towards seed tetrahedra that contain the fiber surface, thereby reducing the search space. This approach also leads to effective analysis of the bivariate field by supporting the identification of relevant fiber surfaces near Jacobi edges.%
} 

\CCScatlist{
    \CCScatTwelve{Human-centered computing}{Visualization}{Visualization techniques}{};
    \CCScatTwelve{Human-centered computing}{Visualization application domains}{Scientific visualization}{};
}

\begin{document}


\firstsection{Introduction}

\maketitle

Data from science and engineering disciplines is often represented as a scalar field over a geometric domain. The scalar field maps each point of the domain to a scalar value. A univariate field refers to a single scalar field and  the term multivariate refers to the case of multiple scalar fields defined over the domain. Several topological structures have been introduced for analyzing a univariate field  and its critical points --  contour tree, Reeb graph, Morse-Smale complex, extremum graph~\cite{heine2016survey}. Extensions to multivariate fields are also known -- Jacobi set~\cite{Edelsbrunner2004jacobiset}, Reeb space~\cite{edelsbrunner2008reeb}, and pareto sets~\cite{huettenberger2014decomposition}. In addition, the notion of fibers and fiber surfaces is introduced as a generalization of isosurfaces to bivariate fields~\cite{carr2015fiber}. A fiber is the preimage of a given pair of scalar values.  A fiber surface is a collection of fibers, the preimage of a line segment or polygon in the range space of the bivariate field.  Visualizing interesting features within multivariate fields in an automated manner and fast computation of these structures has remained a challenge. Extensive work has been done towards the identification of interesting isovalues and for computing the corresponding isosurfaces efficiently. Contour trees and Reeb graphs help locate the seed cells for a particular isosurface, thereby enabling their efficient extraction. The contour tree also serves as a guide for identifying interesting isovalues. However, there is limited work on counterparts for performing similar tasks for bivariate fields, namely for identifying fiber surfaces of interest.  

Our work targets two problems: (a)~Computing fiber surfaces in an output sensitive manner, and (b)~Identifying and computing interesting fiber surfaces in the vicinity of a Jacobi edge. These problems are inspired by the univariate counterparts, where critical point information and contour trees are used to identify isosurface components of interest and to compute them with correctness guarantees.

\myparagraph{Related work.}
Methods and approaches for computing isosurfaces have been extensively studied during the past three decades~\cite{newman2006survey, lorensen2020history}. 
Oostrum \etal~\cite{Oostrum1999contourTreeIE} introduce a seed set based approach for isosurface extraction, that provides significant advantages over the domain and range search based methods that were proposed earlier. The contour tree is an abstract representation of the connectivity of isosurfaces,  a topological structure that capture the connected components of isosurfaces. It also helps locate critical points whose value is close to an input or query isovalue~\cite{heine2016survey}. The located critical points facilitate faster traversal to the isosurface seed cell. Sharma \etal~\cite{sharma2018onDemand} describe an application of augmented contour trees that helps fast location of a seed cell. In this paper, we extend the seed set based approach to bivariate fields towards the computation of a generalization of the isosurface. The Jacobi set~\cite{Edelsbrunner2004jacobiset} is a bivariate counterpart of critical points. Edelsbrunner \etal~\cite{Edelsbrunner2004jacobiset} show the applicability of Jacobi sets to the study of protein interactions and Lagrange points in the solar system. The Jacobi set is typically a large and complex network consisting of multiple edges, and it is difficult to identify important Jacobi edges. Multiple algorithms are available for simplifying the Jacobi set~\cite{Nagaraj2011jacobisetsimplification, bhatia2015localjacobisetsimplification}. 

Carr \etal~\cite{carr2015fiber} generalize the notion of isosurfaces to bivariate fields and introduce fiber surfaces. They demonstrate the use of fiber surfaces to  chemistry and medical imaging data. Carr \etal~\cite{carr2015fiber} use the continuous scatterplots~\cite{Bachthaler2008CSP}, generalization of discrete scatter plots~\cite{sarikaya2017scatterplots}, as an intermediate tool to find  bivariate range values that may represent interesting fiber surfaces. The fiber is defined as a generalization of an isosurface. A collection of fibers result in a fiber surface. Jankowai \etal~\cite{Jankowai2020FeatureLevelSets} introduce feature level-sets, which generalizes isosurfaces and fiber surfaces. Sane \etal~\cite{saneUncertainMultiData2021} extend the idea of univariate confidence isosurfaces to multivariate feature level-sets. Klacansky \etal~\cite{klacansky2016fast} present an algorithm for fiber surface computation with correctness guarantees and improved speed. Various applications~\cite{Blecha2019Nuclear,Raith2019Tensor} have used the fiber surface to study and extract interesting features in multifields. Tierny \etal~\cite{tierny2016jacobi} use Jacobi fiber surfaces for bivariate Reeb space computation and show the application of Reeb space based domain segmentation to peel continuous scatterplot layers. Sakurai \etal~\cite{sakurai2020flexibleFS} introduce an approach towards flexible fiber surface extraction that does not require computation of the Reeb space. However, the method is limited to the exploration of fiber surfaces within the vicinity of a given fiber surface. In this paper, we introduce a method for flexible computation of individual fiber surface components and hence a flexible exploration of the fiber surface while also not requiring expensive computation of the Reeb space. 

\myparagraph{Contributions.}
We introduce a novel output sensitive approach for computation of fiber surfaces for bivariate fields defined on tetrahedral meshes. We present an approach for fast identification of tetrahedra that contain the fiber surface and utilize existing implementation for  extracting the fiber surface within each such tetrahedron. The method works by reducing the search space using the Jacobi set of the bivariate field. Key contributions of our work include
\begin{enumerate}
    \item An output sensitive algorithm for fiber surface computation in tetrahedral meshes using the Jacobi set.
    \item An approach for flexible computation of individual fiber surface components.
    \item Interactive guided exploration of fiber surfaces within the vicinity of Jacobi edges.
\end{enumerate}
\begin{figure*}[!t]
    \centering
    \includegraphics[width=\textwidth]{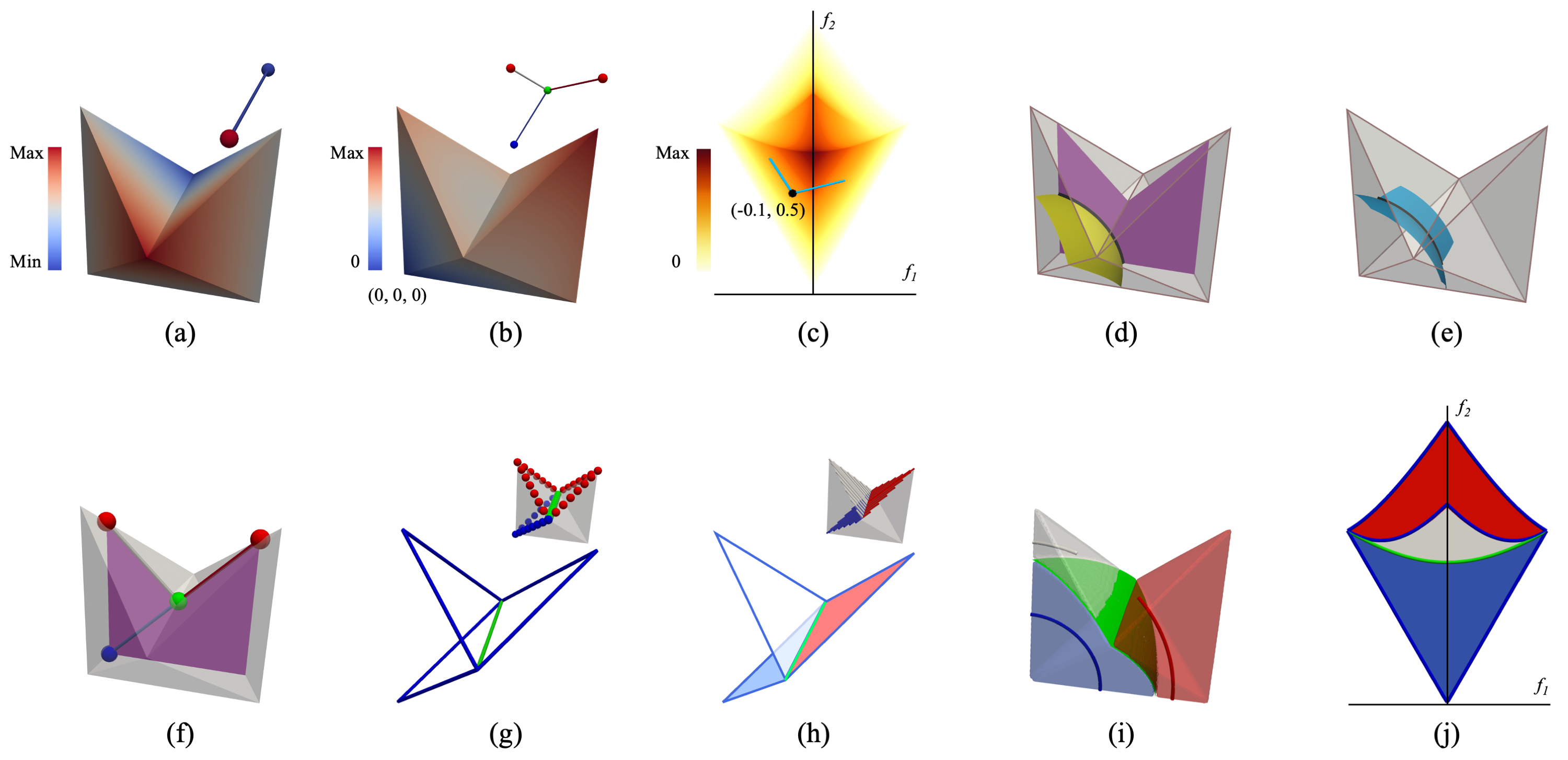}
    \caption{(a,b)~A synthetic bivariate field. (a)~ $f_1$ is the z-coordinate. (b)~$f_2$ is a distance field, the distance of a point from $(0, 0, 0)$. (c)~Continuous Scatter Plot (CSP) of the bivariate field. A blue open control polygon is selected. One of its vertices $(-0.1, 0.5)$  is shown in black. (d)~The purple isosurface of $f_1$ corresponds to isovalue $-0.1$ and the yellow isosurface of $f_2$ corresponds to isovalue $0.5$. The two isosurfaces intersect along the black curve, the fiber corresponding to $(-0.1, 0.5)$. (e)~The blue fiber surface corresponding to the control polygon passing through the black fiber. (f)~Contour tree of $f_2$ for level set (purple) $f_1 = -0.1$ (g)~Jacobi set of the bivariate field. Extremum edges shown in blue and saddle edge shown in green, inset shows stacked critical points of $f_2$ for finite level sets of $f_1$. (h)~Reeb space with different sheets shown in different colors, inset shows the stacked contour trees (i)~Reeb space based domain segmentation, each segment represents same fiber topology (J)~Segmentation projected onto the 2D range space. All CSPs in this paper use a yellow-red color map (\includegraphics[width=0.07\textwidth]{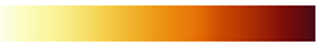}) in log scale to visualize density.} 
    \vspace{-1.2em}
    \label{fig:background}
\end{figure*}

\section{Background}
\label{sec:background}
This section presents a brief introduction to topological analysis of bivariate fields and the mathematical preliminaries necessary for describing our algorithm. For a detailed description of Morse theory, topological descriptors for scalar fields and the relevant definitions, we refer the reader to surveys and books on the topic~\cite{heine2016survey, yan2021scalar}.

\myparagraph{Univariate field.} A \emph{scalar field}, also called a \emph{univariate field}, maps each point of a spatial domain $\mathcal{M}$ to a single scalar value.
\begin{align*}
f_u : \mathcal{M} \rightarrow \mathbb{R}.
\end{align*}
Generally, the scalar value represents a physical quantity like temperature, pressure, speed, height, or distance. Non-physical quantities such as probability distributions may also be represented as a scalar field.  \autoref{fig:background}(a) shows a univariate field where each domain point is mapped to z-coordinate and visualized using a color map.

\myparagraph{Isosurface.} Given a scalar value $a$, the \emph{isosurface} $\mathcal{I}$ represents all points in $\mathcal{M}$ that map to $a$, 
\begin{align*}
\mathcal{I} = f_u^{-1} (a).
\end{align*}
The scalar $a$ is referred to as an isovalue.

\myparagraph{Bivariate field.} A collection of scalar fields defined over a common domain is called as \emph{multivariate field} or \emph{multifield}. A \emph{bivariate field} is a special instance of a multifield when two scalar fields are defined over $\mathcal{M}$,
\begin{align*}
f = \{f_1,f_2\} : \mathcal{M} \rightarrow \mathbb{R}^2.
\end{align*}
Analyzing a bivariate field helps explore the relationship between the individual scalar fields $f_1$ and $f_2$ and to study the features impacted by both fields simultaneously. For example, electron density field and its gradient magnitude may help in the extraction of atoms or bonds in a molecule. \autoref{fig:background}(a) and \autoref{fig:background}(b) show a synthetic dataset~\cite{Tierny2018ttk}. Each point of the spatial domain is mapped to a bivariate field $\{f_1, f_2\}$ where $f_1$ is the z-coordinate and $f_2$ is the distance of each point from $\{0, 0, 0\}$.

\myparagraph{Continuous scatterplot.} A CSP is a generalization of the discrete scatterplot to spatially continuous multifields~\cite{Bachthaler2008CSP}. In this paper, we only consider bivariate fields. The CSP maps a point from the 2D range space of the bivariate field to the density of that point. Density is the continuous counterpart of frequency of data points in a discrete scatterplot. A point in the CSP with high density implies a high number of occurrence for the corresponding pair of values $(f_1 = s_1,f_2 = s_2)$. \autoref{fig:background}(c) shows the CSP of a bivariate field. Visual inspection of the CSP helps identify interesting values in the range space, say high or low density values or a unique pattern.

\myparagraph{Fiber.} A fiber~\cite{carr2015fiber} is the multivariate counterpart of an isosurface. Given a bivariate isovalue $(s_1,s_2)$, a \emph{fiber} $\mathcal{F}$  is the collection of points in  $\mathcal{M}$ that map to $(s_1,s_2)$ under $f$, 
\begin{align*}
\mathcal{F} = f^{-1}(s_1, s_2)
\end{align*}
\autoref{fig:background}(d) shows the isosurface $f_1^{-1}(s_1)$ in purple and $f_2^{-1}(s_2)$ in yellow. The intersection of these two isosurfaces shown in black, is the fiber $\mathcal{F}$. $\mathcal{F}$ is the preimage of the black point shown in \autoref{fig:background}(c).

\myparagraph{Fiber surface.} The preimage of a collection of points that lie along a continuous curve in the CSP is called a \emph{fiber surface}. So, the fiber surface is a special collection of fibers. The blue fiber surface in \autoref{fig:background}(e) corresponds to the 2-edge polygon (blue) in \autoref{fig:background}(c). The black fiber belongs to this fiber surface.

\myparagraph{Control polygon.} The set of two blue edges shown in \autoref{fig:background}(c) is called a control polygon. A \emph{control polygon} is a polygon embedded in the range space of the bivariate field $f$. It may be open or closed and serves as an interface to specify interesting fiber surface in the spatial domain. The topology of the fiber surface depends on the control polygon. A closed control polygon corresponds to a closed fiber surface unless it intersects the domain boundary.
\begin{figure}[!t]
    \centering
    \includegraphics[width=0.3\textwidth]{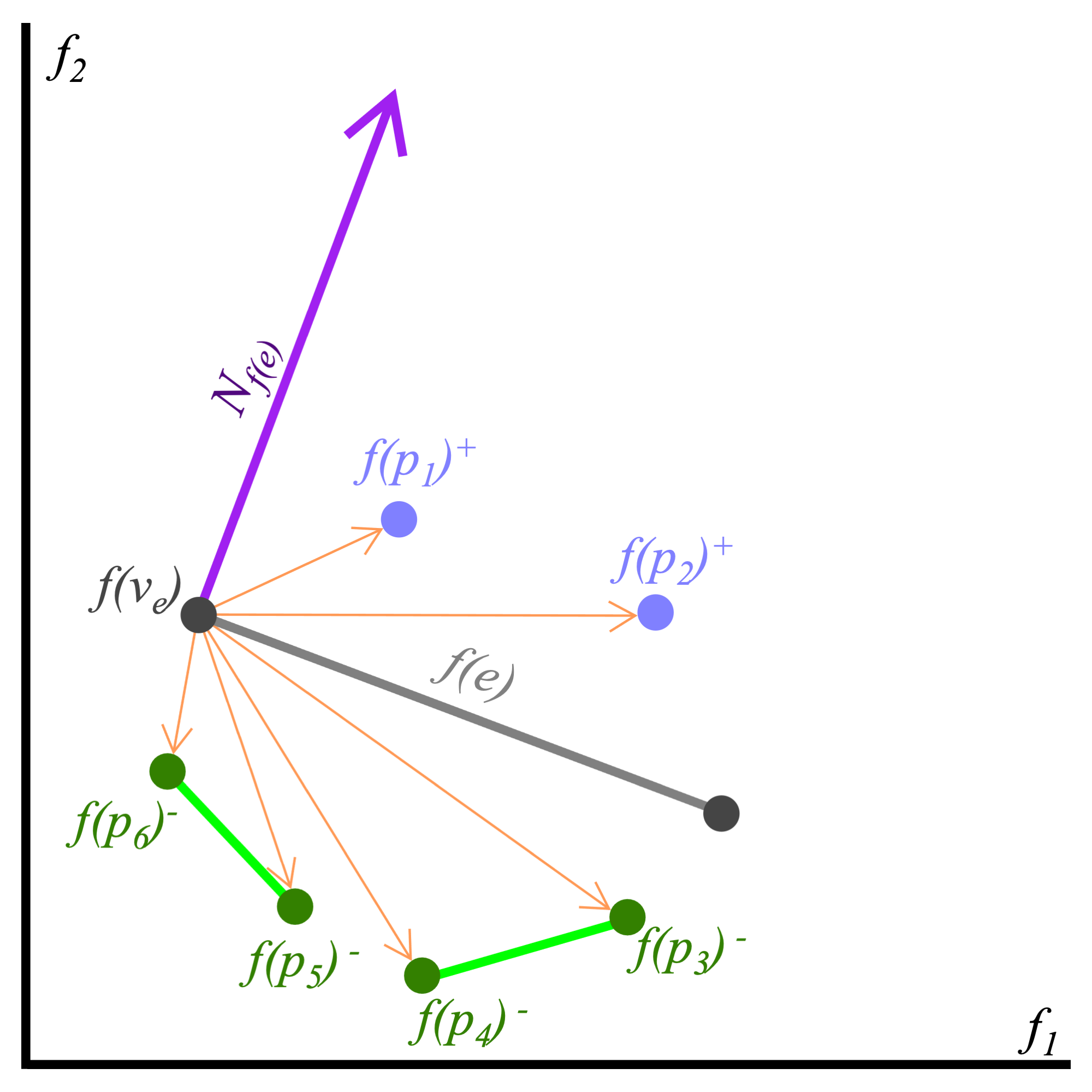}
    \caption{Classifying a Jacobi edge based on $Lk^-(e)$ and $Lk^+(e)$. The green points are identified in $Lk^-(e)$ and blue in $Lk^+(e)$. Both $Lk^+(e)$ and $Lk^-(e)$ have two disconnected components and hence the edge is classified as a saddle Jacobi edge.} 
    \vspace{-1.2em}
    \label{fig:jacobiEdgeClassification}
\end{figure}

\begin{figure}[!t]
    \centering
    \includegraphics[width=0.5\textwidth]{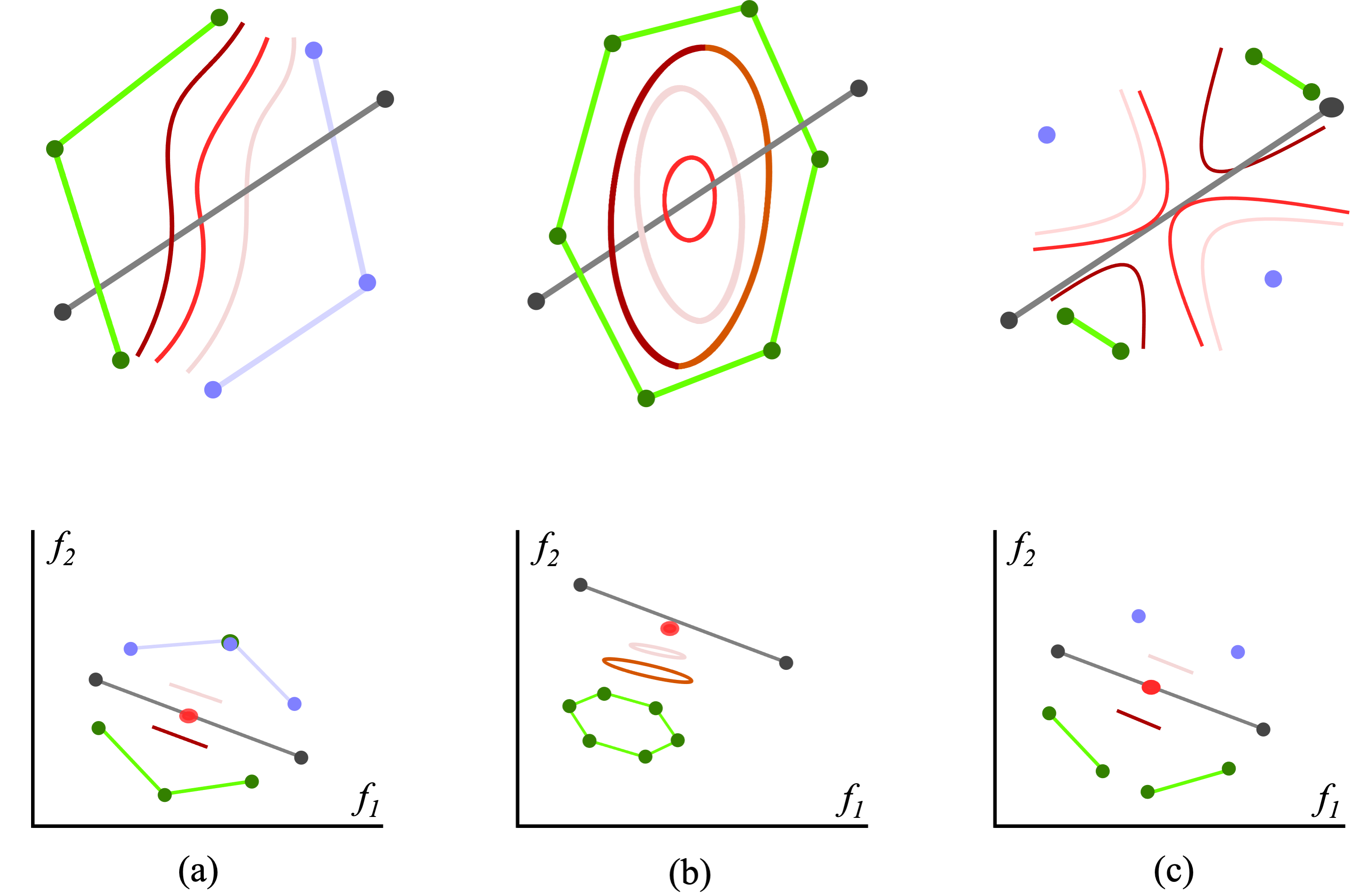}
    \caption{(Top)~Fiber topology in the neighborhood of an edge (black) in the domain. (Bottom)~Corresponding projections in range space. The lower link is represented in green and upper link in blue. (a)~Fiber topology does not change upon crossing a regular edge in the range space. (b)~Fibers originate or vanish at an extremum Jacobi edge. (c)~Fibers split or merge at a saddle Jacobi edge.} 
    \vspace{-1.2em}
    \label{fig:jacobiEdgeTypes}
\end{figure}

\myparagraph{Jacobi set.}  The Jacobi set~\cite{Edelsbrunner2004jacobiset} of two Morse functions defined on a d-manifold is the set of critical points of the restrictions of one function to the isosurfaces of the second function. It can also be considered as the set of points where the gradients of both functions are parallel or one of the gradients vanish. Let $\Delta f_1(x)$ and $\Delta f_2(x)$ denote the gradients of scalar fields $f_1$ and $f_2$, then the Jacobi set is defined as
\begin{align*}
\mathcal{J} = \{x\in M\ |\ \Delta f_1(x) +\lambda \Delta f_2(x) = 0\ or \\
\lambda \Delta f_1(x) + \Delta f_2(x)= 0\}
\end{align*}
We follow Klacansky \etal~\cite{klacansky2016fast} to present an illustration of the Jacobi set using a simple example.
\autoref{fig:background}(f) shows the contour tree of $f_2$ restricted to an isosurface of $f_1$. If contour trees for all isosurfaces of $f_1$ are stacked together then the set of critical points of the restrictions of $f_2$ form individual edges of the Jacobi set. A Jacobi edge can be considered as a bivariate analog of the critical point. In a univariate setting, the topology of isosurfaces change in the vicinity of critical points. Similarly, topology of fibers change in the vicinity of Jacobi edges. A Jacobi edge can be categorized as a saddle or extremum edge. The categorization is based on the connectedness of its lower and upper links. The star $st(s)$ of a simplex $s$ is defined as the set of simplices that have $s$ as a face. Link $Lk(s)$ of a simplex $s$ is the set of faces of $st(s)$ that do not intersect with $s$. In the univariate setting, a critical point partitions the 1D range space into two halves, smaller scalar values go to the lower half and higher value to the upper. Lower and upper links are subsets of the link that correspond to this partition. In the bivariate scenario, a Jacobi edge also partitions the 2D range space in two halves. Any point from the domain fall on one side of the edge as shown in \autoref{fig:jacobiEdgeClassification}.  Simulation of Simplicity~\cite{Edelsbrunner1994SoS}  is employed to ensure that no two edges $e_1$ and $e_2$  have collinear images  and no edge can have end points mapped to the same bivariate value.  In order to categorize an edge $e$ as Jacobi edge, a signed distance measure is defined for each point $p$
\begin{align*}
d_s(p) = \overrightarrow{f(p)f(v_e)}.\overrightarrow{N_{f(e)}}.
\end{align*}
$d_s$ is the dot product of two vectors -- the normal vector $N_{f(e)}$ to the image $f(e)$ of edge $e$ and the vector connecting an end point $f(v_e)$ of $f(e)$ to $f(p)$. The dot product $d_s$ is positive for points lying on one side of $f(e)$ and negative for points lying on the other side. A point $p$ is categorized in lower link $Lk^-(e)$ if $d_s(p) < 0$ else in $Lk^+(e)$. The edge $e$ is classified as an extremum Jacobi edge if either of $Lk^-(e)$ or $Lk^+(e)$ is empty, a saddle Jacobi edge if any of $Lk^-(e)$ or $Lk^+(e)$ consists of more than one connected component, else as a regular edge. In \autoref{fig:jacobiEdgeClassification}, $p_1$ and $p_2$ are identified as $Lk^+(e)$; $p_3$, $p_4$, $p_5$ and $p_6$ as $Lk^-(e)$. Both the lower and upper link have two disconnected components and hence $e$ is identified as the saddle edge. The topology of fibers in the vicinity of Jacobi edges in the range space changes in the similar manner. Topology stays the same upon crossing a regular edge, refer \autoref{fig:jacobiEdgeTypes}(a). The fibers originate or vanish around an extremum edge, refer \autoref{fig:jacobiEdgeTypes}(b). Fibers split or merge in the neighborhood of a saddle edge, refer \autoref{fig:jacobiEdgeTypes}(c). \autoref{fig:background}(g) shows the computed Jacobi edges, the saddle in green and extremum edges in blue, the inset shows a finite number of stacked critical points of $f_2$ on different level sets of $f_1$.

\myparagraph{Reeb space.} Reeb space is the bivariate analog of the Reeb graph. In univariate scenario, if a range of scalar values is swept from minimum to maximum and each level set component is contracted to a point, the resulting structure is called the Reeb graph. For the bivariate counterpart, if each fiber component is contracted to a point, the resulting structure is called the Reeb space. Alternatively, it can be computed by stacking the Reeb graphs of one scalar field restricted to the infinite family of isosurfaces of the second function. \autoref{fig:background}(h) shows the Reeb graph computed for a synthetic bivariate field. Inset shows the stacked Reeb graphs of $f_2$ computed over a finite number of isosurfaces of $f_1$. Boundary of each arc in the Reeb graph are the critical points. The arc forms a surface in the Reeb graph and critical points form the boundary of that surface called Jacobi edges. We refer these surfaces as ``sheets'' in rest of the discussion. Each sheet of the Reeb graph represents fibers of same topology lying inside corresponding 3D segment in the domain. The synthetic bivariate field has three different segments shown as blue, red, and white in \autoref{fig:background}(i) corresponding to three surfaces of Reeb graph bounded by Jacobi edges. 

\myparagraph{Jacobi fiber surfaces.} Image $f(e)$ of a Jacobi edge $e$ is an edge in the range space. If $f(e)$ is used as a control polygon, the resulting fiber surface passes through $e$ and is called a Jacobi fiber surface. Saddle fiber surface correspond to saddle Jacobi edge and extremum fiber surface correspond to extremum Jacobi edge. Topology of fibers change only in the vicinity of a Jacobi fiber surface. A fiber can have multiple components similar to an isosurface. A fiber corresponding to a point lying in the vicinity of $f(s)$, where $s$ is the saddle Jacobi edge, may have more than one component. Jacobi fiber surfaces partition the domain into 3D segments. Each segment is a collection of fibers with the same topology. In \autoref{fig:background}(i), the blue fiber in the blue segment corresponds to bivariate value $\{0, 0.4\}$. As $f_2$ increases for a fixed value of $f_1=0$, the blue fiber remains a single component until it crosses the green fiber surface when it splits into two components within the red and white domain segments. The map of each 3D segment in the range space is bounded by the corresponding Jacobi edges. In \autoref{fig:background}(j), the green saddle Jacobi edge separates the blue segment from the other segments. White and red segments are also bounded by corresponding Jacobi edges but overlap in the range space due to overlapping bivariate values.

\section{Jacobi set based fiber surface extraction}
\label{sec:method}
We now describe an algorithm for computing the fiber surface using an efficient directed search to locate seed tetrahedra followed by a surface traversal step. The algorithm is designed based on the intuition gained from the univariate counterpart. In the case of a univariate field, the augmented contour tree is queried to gain direct access to seed tetrahedra that correspond to an isovalue~\cite{carr2010flexible}. A direct extension to bivariate fields would use the Reeb space to identify seeds. Given the computational challenges in constructing and storing the Reeb space, we propose the use of the simpler Jacobi set. The Jacobi set ($\mathcal{J}$) of a bivariate field $f$ defined over a tetrahedral mesh ($\mathcal{M}$) and a control polygon ($\mathcal{C}$) defined on the range space of $f$ are given as input. $\mathcal{J}$ remains the same for a particular bivariate field, hence it can be precomputed and reused for different control polygons. The Jacobi set is computed using TTK~\cite{Tierny2018ttk}. Our algorithm locates and extracts all tetrahedra that contain the fiber surface corresponding to an edge $(u,v)$ of $\mathcal{C}$. The fiber surface computation for each control polygon edge can be executed independently. We now describe a four-step algorithm that computes the fiber surface for a control polygon edge $(u,v)$. This algorithm can be iteratively executed for each edge of $\mathcal{C}$. Algorithm~\ref{alg:algo} shows pseudo code for the three steps that locate all tetrahedra that contain the fiber surface.

\begin{figure}[!t]
    \centering
    \includegraphics[width=0.3\textwidth]{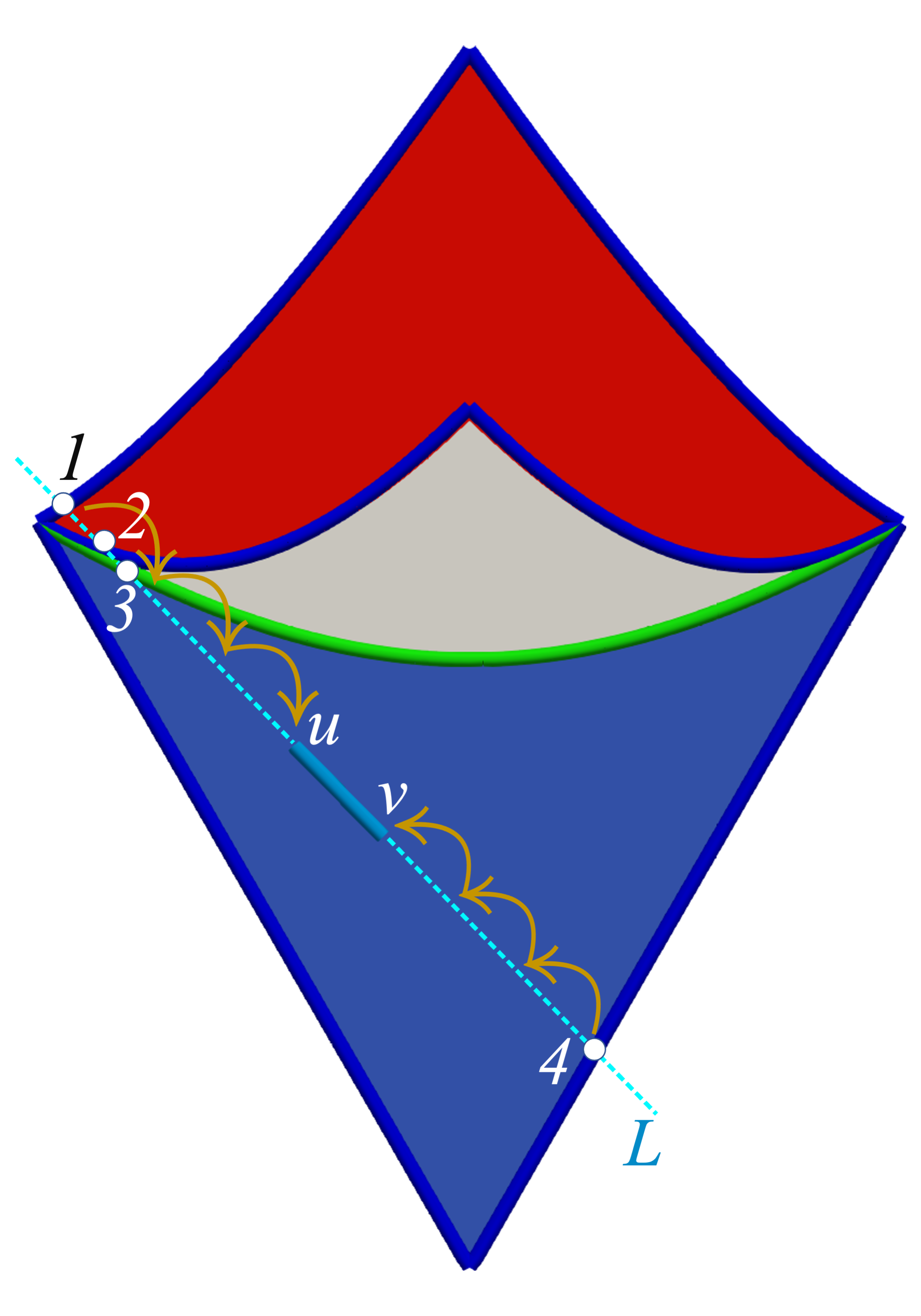}
    \caption{Control edge $(u,v)$ overlapping with blue Reeb space sheet and enclosed by intersections with Jacobi edges. Line $L$ (dashed) extended through $(u,v)$ (solid) intersects with four Jacobi edges. Directed search is initiated from all, to find the seed cell overlapping with either $u$ or $v$.} 
    \vspace{-1.2em}
    \label{fig:directedSearch}
\end{figure}

\myparagraph{(A) Jacobi intersections.} In the univariate setting, the input isovalue is used to traverse the contour tree and identify the arcs that span the isovalue. A directed search to find the seed cell is initiated from one end point of that arc, both of which are critical points. In the bivariate scenario, the aim is to locate the Reeb space sheet(s) containing $(u,v)$ and extract the set ($J_s$) of Jacobi edges bounding the sheet(s). Any edge from $J_s$ may be used to initiate the directed search. The Reeb space is not available to our algorithm. Instead, we compute the Jacobi edges that intersect the line $L$ containing the edge $(u,v)$, see \autoref{fig:directedSearch}. The set of intersected Jacobi edges $J_{int}$ contains $J_s$ together with additional edges. In \autoref{fig:directedSearch}, edge $3$ or edge $4$ would yield the required seed cell since $(u,v)$ lies inside the blue segment, but edges $1$ and $2$ are also identified as intersected Jacobi edges. Line~3 in Algorithm~\ref{alg:algo} implements this step. The running time of this Jacobi edge intersection computation step is  $O(|\mathcal{J}|)$.

\myparagraph{(B) Directed search.} Each intersected Jacobi edge may or may not yield a seed tetrahedron. In order to find out the seed tetrahedron, a directed breadth first search (BFS) is initiated from the cell (tetrahedron) containing the Jacobi edge. Lines~6 to~14 in Algorithm~\ref{alg:algo} show the pseudo code for this directed search. In each iteration of the directed BFS, a neighbor tetrahedron whose edge intersects  $L$ and has the closest intersection point to either endpoint of  $(u,v)$, is added to the BFS queue (Line~14 in Algorithm~\ref{alg:algo}). The aim is to select a neighbor tetrahedron that helps to move towards the seed cell intersecting with $(u,v)$. Each edge $e$ of a tetrahedron maps to an image $f(e)$ in the range space. Intersection of $L$ and $f(e)$ for all edges $e$ of every neighbor tetrahedron is computed. If there is no intersection then clearly the tetrahedron will not lead to $(u,v)$ and can be discarded. Else, the euclidean distance inside range space between the intersection point to $u$ and $v$ is computed. The neighbor tetrahedron with the edge having minimum distance is chosen. The directed search terminates either upon reaching a cell intersected by $(u,v)$ in the range space (Line~9) or when there exist no new cell to explore \ie upon reaching a dead end or boundary point is reached (Line~15). The BFS inherently provides an optimization, namely that no directed search is initiated from a tetrahedron containing an intersected Jacobi edge if it is visited during a directed search initiated elsewhere. The time taken for the directed search depends on the number of visited tetrahedra $|D_t|$. Clearly, $D_t$ cannot exceed the total number of tetrahedra $T$. Hence, the running time is  $O(|D_t|) = O(|T|)$.

\myparagraph{(C) Restricted BFS} Next, we extract the tetrahedra containing the fiber surface by initiating a restricted BFS from each seed cell. The BFS explores only those cells that overlap with $(u,v)$. All cells yielded in this step contain some part of the fiber surface. BFS initiated from each of the seed cells will yield at least one fiber surface component but multiple seeds may belong to the same component. Hence, seeds that are visited during a restricted BFS initiated from another seed cell are discarded. If the required fiber surface passes through $|Z|$ tetrahedra then this step traverses exactly $O(|Z|)$ tetrahedra.

\begin{algorithm*}

\caption{Extract all tetrahedra containing the fiber surface}
\label{alg:algo}
\begin{algorithmic}[1]

\Procedure{extractFiberSurfaceTets}{$\mathcal{M}, f, \mathcal{J}, (u,v)$}\Comment{Returns Z: Tetrahedra containing the fiber surface of $(u,v)$}

    \State $L \gets line\ passing\ through\ (u,v)$
    \State $J_{int} \gets Jacobi\ edges\ intersected\ by\ L$
    \State $Z \gets NULL$

    \ForAll{$j_i\ in\ J_{int}$} \Comment{Initiate directed search from each intersected Jacobi edge}
        \State $start \gets tetrahedron\ containing\ j_i$
        \State $seedT \gets NULL$
            \While {true}
                \If{$start\ contains\ either\ u\ or\ v$}
                    \State $seedT \gets start$ 
                    \State $break$
                \EndIf
                \State $prevStart \gets start$
                \State $start \gets  traverseToNeighborClosestToUV(start)$\Comment{Directed traversal}
                \If{$start\ ==\ prevStart$} \Comment{Dead end}
                    \State $break$
                \EndIf
            \EndWhile
        \State $Z \gets Z\ +\ restrictedBFS(seedT)$
    \EndFor
    \State $\textbf{return}\ Z$
    \EndProcedure
\end{algorithmic}
\end{algorithm*}

\myparagraph{(D) Fiber surface extraction.} The set of tetrahedra ($Z$) extracted in the previous step contain the required fiber surface. We extract these tetrahedra from the domain and supply them as input to TTK, to reuse their implementation for computing the fiber surface within each tetrahedron~\cite{klacansky2016fast}. The running time for this step is linear in the number of input tetrahedra, $O(|Z|)$.

\myparagraph{Correctness.} The restricted BFS  traverses through all cells reachable from a seed cell while ensuring that the range of scalar values within the traversed cell overlap with the control polygon edge $(u,v)$ in the range space.We prove the correctness of the algorithm by showing that it locates at least one seed cell for each connected fiber surface component. If a control polygon edge $(u,v)$ maps to $n$ connected fiber surface components then it will overlap with exactly $n$ Reeb space sheets. In the projected range space, the overlapping segments of $(u,v)$ will be enclosed by bounding Jacobi edges of the corresponding sheets. The infinite line $L$ extended through $(u,v)$ will intersect with at least one of those enclosing Jacobi edges. As discussed in \autoref{sec:background}, each surface in the Reeb space represents fibers having same topology without any split or merge. Hence, a directed search initiated from a valid Jacobi edge of that particular sheet will locate the seed cell.

\myparagraph{Runtime analysis.} Adding the time taken for the four steps, the total running time of the algorithm is $O(|\mathcal{J}| + |D_t| + |Z|) = O(T)$. In practice, the run time crucially depends on the number of Jacobi edges and the number of intersected Jacobi edges.

\section{Results}
\label{sec:results}
We now describe results of experiments conducted on four different datasets: Ethanediol, Thiophene-Quinoxaline, Tooth, and Combustion. First, we study the correctness of the results via comparisons against results obtained using a previous method implemented in TTK~\cite{Tierny2018ttk}. We also compare the runtime performance of our Jacobi set driven search against a previous method that employs domain search~\cite{klacansky2016fast}. Next, we highlight specific applications that are supported by our output sensitive approach to fiber surface extraction.

\subsection{Quality}
\begin{figure*}[!t]
    \centering
    \includegraphics[width=\textwidth]{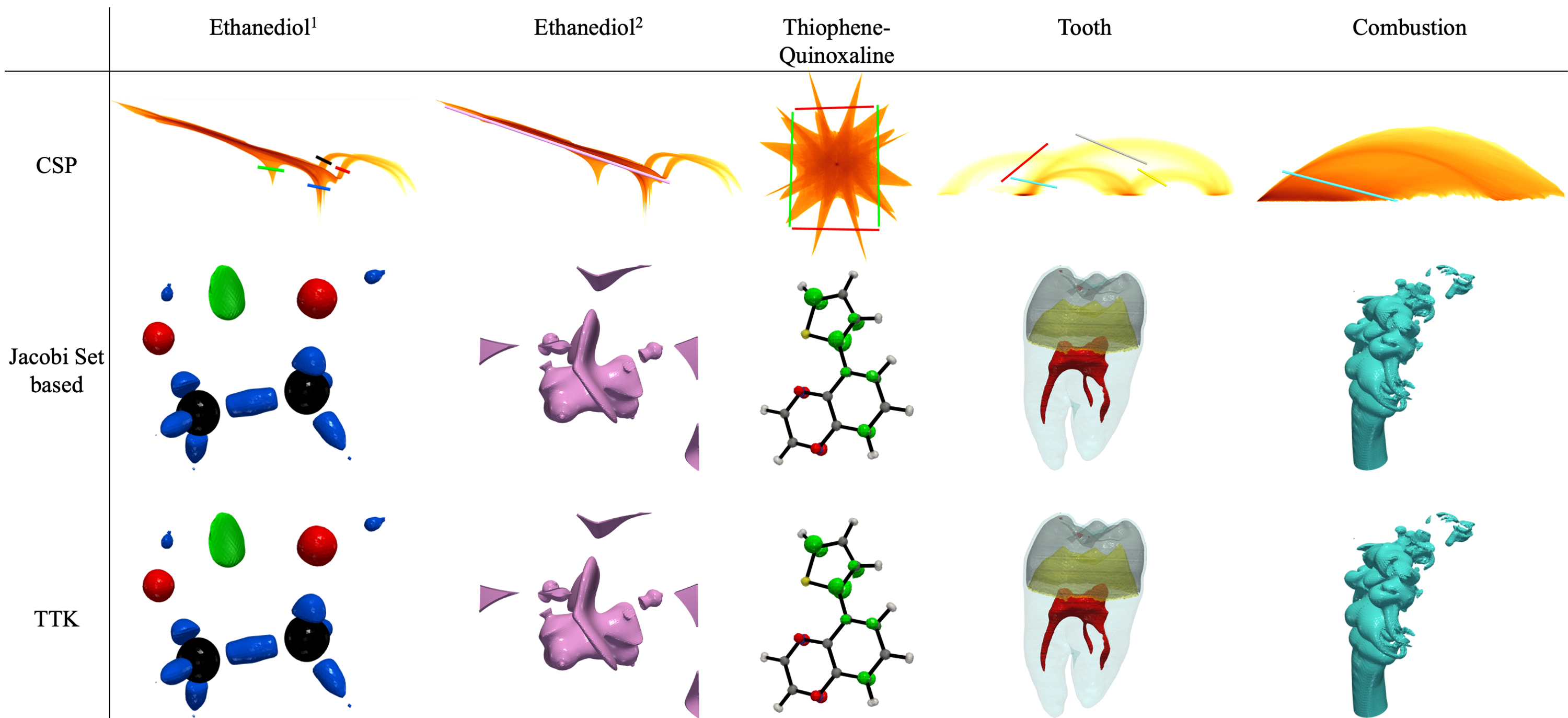}
    \caption{Visual comparison. Fiber surfaces computed by the proposed Jacobi set based search algorithm match with those computed using the fast and exact algorithm implemented in TTK~\cite{klacansky2016fast}.} 
    \vspace{-1.2em}
    \label{fig:applications}
\end{figure*}

\autoref{fig:applications} shows a visual comparison of fiber surfaces generated using our Jacobi set driven search algorithm and the existing algorithm~\cite{klacansky2016fast}, implemented in TTK. We have selected a control polygon containing a single edge for simplicity. The algorithm can be executed independently for each edge of a generic control polygon, as mentioned in \autoref{sec:method}. control polygons and their corresponding fiber surfaces are shown using a common color. Visually, the fiber surfaces computed by our algorithm and TTK look the same. We also compare the expected number of tetrahedra intersected by the fiber surface against the number of tetrahedra reported by our algorithm. The expected number of tetrahedra intersecting with fiber surface is computed by traversing the entire domain. \autoref{table:quality} shows the corresponding results. We observe an exact match, signifying that no tetrahedron containing the fiber surface is missed by our algorithm.
\begin{figure*}[!t]
    \centering
    \includegraphics[width=\textwidth]{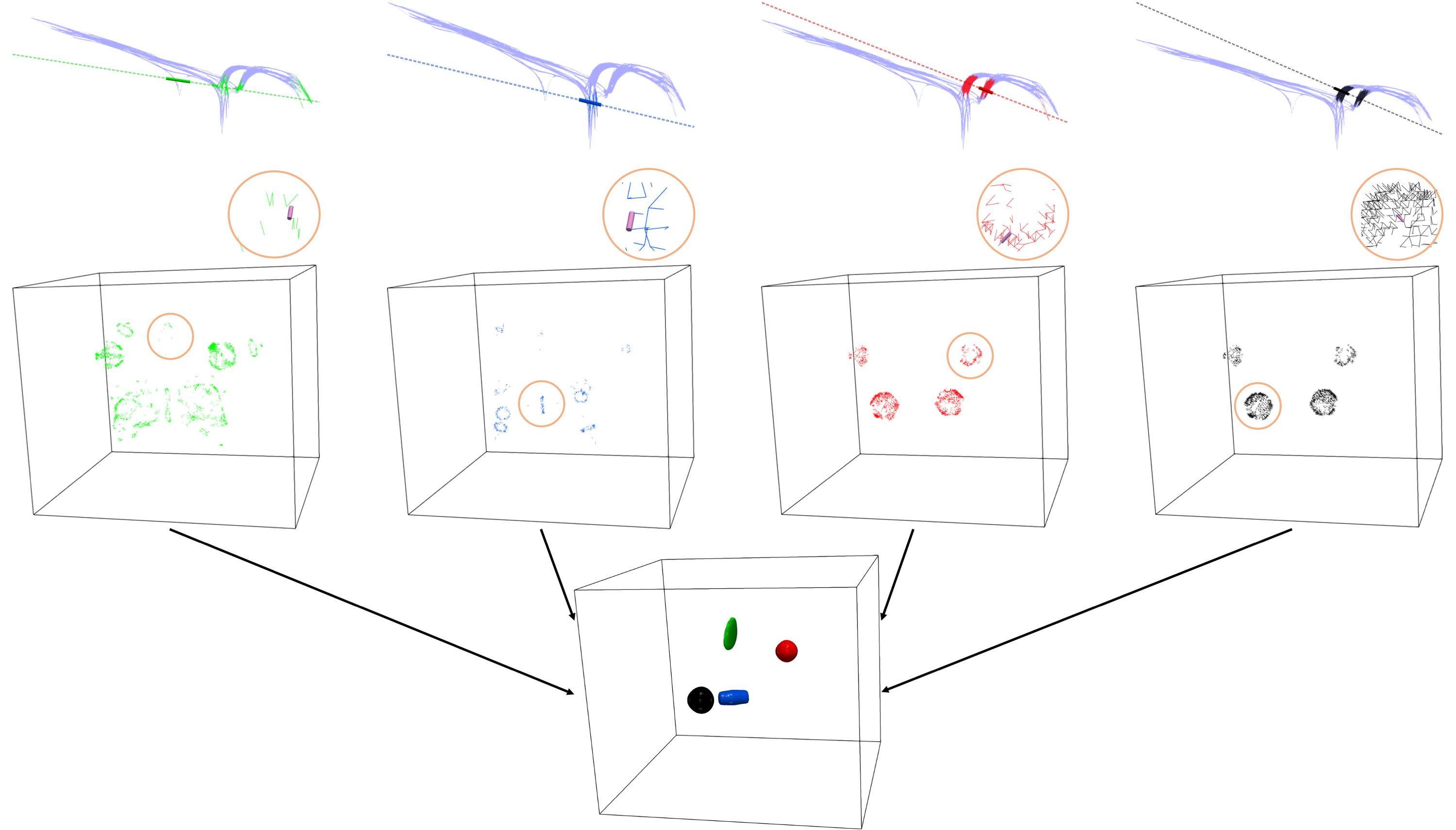}
    \caption{Flexible fiber surfaces extracted from the Ethanediol dataset. (Top)~Jacobi edges in the range space. Intersected Jacobi edges are highlighted in corresponding color. (Middle)~Intersected Jacobi edges are shown in domain space to facilitate interactive selection of an edge, with the objective of exploring fiber surface in its neighborhood. Inset shows the selected edge in pink. (Bottom)~Fiber surface components extracted for the selected Jacobi edges.} 
    \vspace{-1.2em}
    \label{fig:flexibleEthane}
\end{figure*}
\begin{figure*}[!t]
    \centering
    \includegraphics[width=0.7\textwidth]{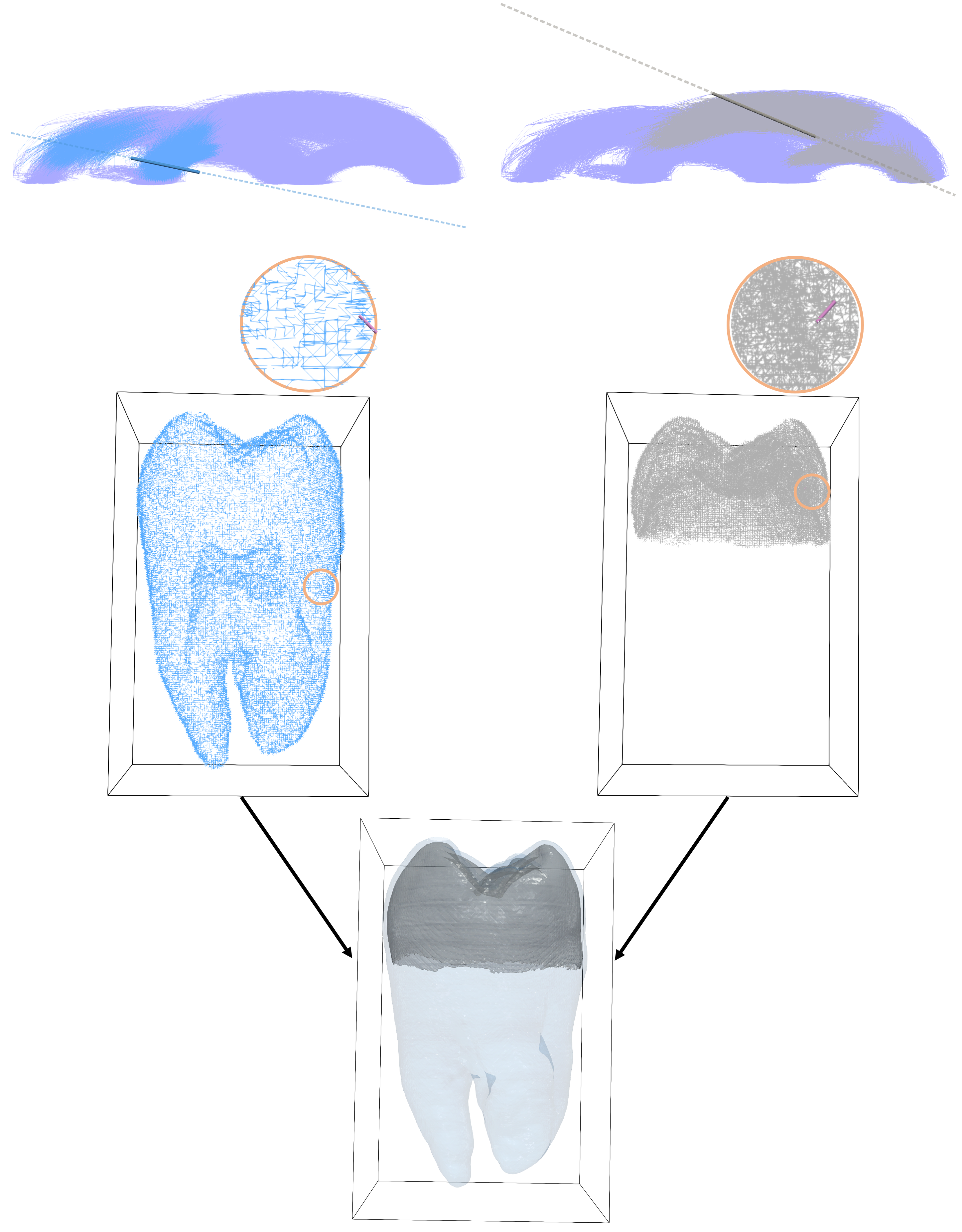}
    \caption{Flexible fiber surfaces extracted from the Tooth dataset. (Top)~Jacobi edges in the range space. Intersected Jacobi edges are highlighted in corresponding color. (Middle)~Intersected Jacobi edges shown in domain space to facilitate interactive selection of an edge, with the objective of exploring fiber surface in its neighborhood. Inset shows the selected edge in pink. (Bottom)~Fiber surface components extracted for the selected Jacobi edges.} 
    \vspace{-1.2em}
    \label{fig:flexibleTooth}
\end{figure*}
\begin{table}[!t]
    \centering
\begin{tabular}{ p{1.5cm}p{1.2cm}>{\raggedleft\arraybackslash}p{1.5cm}>{\raggedleft\arraybackslash}p{2cm} } 
\small{Dataset} & \small{Control polygon} & \small{\#Tets containing FS} & \small{\#Tets using Jacobi set driven search}\\
\hline
\multirow{4}{4em}{Ethanediol$^1$} & Blue & 12804 & 12804\\ 
& Green & 3524 & 3524\\
& Red & 4718 & 4718\\ 
& Black & 7612 & 7612\\ 
\hline
Ethanediol$^2$ & Pink & 138239 & 138239\\
\hline
\multirow{2}{4em}{Thiophene-Quinoxaline} & Green & 4032 & 4032\\ 
& Red & 946 & 946\\
\hline
\multirow{4}{4em}{Tooth} & Blue & 223145 & 223145\\ 
& Yellow & 47796 & 47796\\ 
& Red & 27645 & 27645\\
& Grey & 87185 & 87185\\ 
\hline
Combustion & Blue & 157287 & 157287\\ \\

\end{tabular}
\caption{Validating the algorithm by comparing the number of tetrahedra containing the fiber surface (FS) against an exhaustive search.}
\label{table:quality}
\end{table}

\begin{table*}[!t]
    \centering
\begin{tabular}{p{1.5cm}p{1cm}p{1cm}p{1cm}>{\raggedleft\arraybackslash}p{1.5cm}>{\raggedleft\arraybackslash}p{1.5cm}||>{\raggedleft\arraybackslash}p{1.25cm}>{\raggedleft\arraybackslash}p{1.25cm}>{\raggedleft\arraybackslash}p{1cm}>{\raggedleft\arraybackslash}p{1.5cm}>{\raggedleft\arraybackslash}p{1cm}}
&&&&&& \multicolumn{4}{c}{\small{Computation times (msec)}}\\
\hline
 \small{Dataset} & \small{\#Cells} & \small{\#Jacobi edges} & \small{Control polygon} & \small{\#Jacobi intersections} & \small{\#Tets containing FS} & \small{Jacobi intersections (A)} & \small{Directed Search (B)} & \small{Restricted BFS (C)} & \small{Traversal (T = A + B + C)} & \small{Data transfer}\\
\hline
\multirow{4}{4em}{Ethanediol$^1$} & \multirow{4}{4em}{8718150} & \multirow{4}{4em}{95935} & Blue & 394 & 12804 & 31 & $0^*$ & 57 & 88 & 454\\ 
& & & Green & 2066 & 3524 & 31 & 23 & 15 & 69 & 446\\ 
& & & Red & 1318 & 4718 & 31 & 11 & 21 & 63 & 448\\
& & & Black & 1556 & 7612 & 30 & 5 & 33 & 68 & 446
\\  
\hline
Ethanediol$^2$ & 8718150 & 95935 & Pink & 1048 & 138239 & 31 & 1 & 629 & 661 & 565\\

\hline
\multirow{2}{4em}{Thiophene-Quinoxaline} & \multirow{2}{4em}{2479680} & \multirow{2}{4em}{20737} & Green & 38 & 4032 & 17  & $0^*$  & 18 & 35 & 258
\\ 
& & & Red & 26 & 946 & 14 & $0^*$ & 4 & 18 & 252
\\ 
\hline
\multirow{4}{4em}{Tooth} & \multirow{4}{4em}{7588800} & \multirow{4}{4em}{2732204} & Blue & 44821 & 223145 & 897 & 95 & 1021 & 2013 & 586
\\ 
& & & Yellow & 39185 & 47796 & 874 & 273 & 217 & 1364 & 428
\\ 
& & & Red & 7236 & 27645 & 879 & 11 & 123 & 1013 & 406
\\
& & & Grey & 57149 & 87185 & 885 & 351 & 397 & 1633 & 463
\\  
\hline
Combustion & 18675345 & 2449505 & Blue & 19668 & 157287 & 819 & 30 & 757 & 1606 & 1093\\ \\

\end{tabular}
\caption{Time taken by Jacobi set driven search algorithm to locate all tetrahedra containing the required fiber surface (FS) and time to transfer to TTK for fiber surface computation. Time taken to compute the Jacobi intersections is directly proportional to number of Jacobi edges. Time for executing the restricted BFS increases with number of tetrahedra that contain the fiber surface. Ethanediol$^1$ and Ethanediol$^2$ are identical datasets; the control polygon chosen in Ethanediol$^2$ results in a larger sized fiber surface. $0^*$ represents values smaller than 1 msec.}
\label{table:traversal}
\end{table*}

\begin{table*}[!t]
    \centering
\begin{tabular}{p{1.5cm}p{1cm}p{1.75cm}>{\raggedleft\arraybackslash}p{2cm}||>{\raggedleft\arraybackslash}p{1.75cm}>{\raggedleft\arraybackslash}p{1.75cm}>{\raggedleft\arraybackslash}p{1.5cm}||>{\raggedleft\arraybackslash}p{1.75cm}>{\raggedleft\arraybackslash}p{1.25cm}} 
&&&& \multicolumn{3}{c}{\small{Using Jacobi set (msec)}} & \multicolumn{2}{c}{\small{TTK (msec)}}\\
\hline
\small{Dataset} & \small{\#Cells} & \small{Control polygon} & \small{\#Tets containing FS} & \small{Traversal (T)} & \small{FS Extraction (D)} & \small{Total (T+D)} & \small{Without octree} & \small{With octree}
\\
\hline
\multirow{4}{4em}{Ethanediol$^1$} & \multirow{4}{4em}{8718150} & Blue & 12804 & 88 & 21 & 109 & 472 & 28\\
& & Green & 3524 & 69 & 5 & 74 & 472 & 10\\ 
& & Red & 4718 & 63 & 8 & 71 & 461 & 11\\ 
& & Black & 7612 & 68 & 13 & 81 & 471 & 18
\\  
\hline
Ethanediol$^2$ & 8718150 & Pink & 138239 & 661 & 237 & 898 & 675 & 726\\

\hline
\multirow{2}{4em}{Thiophene-Quinoxaline} & \multirow{2}{4em}{2479680} & Green & 4032 & 35  & 6 & 41 & 259 & 8
\\ 
& & Red & 946 & 18 & 1 & 19 & 257 & 2
\\ 
\hline
\multirow{4}{4em}{Tooth} & \multirow{4}{4em}{7588800} & Blue & 223145 & 2013 & 386 & 2399 & 767 & 519
\\ 
& & Yellow & 47796 & 1364 & 83 & 1447 & 485 & 113
\\ 
& & Red & 27645 & 1013 & 47 & 1060 & 424 & 81
\\ 
& & Grey & 87185 & 1633 & 151 & 1784 & 579 & 195
\\  
\hline
Combustion & 18675345 & Blue & 157287 & 1606 & 271 & 1877 & 1211 & 339\\ \\

\end{tabular}
\caption{Total time taken by our algorithm for fiber surface (FS) computation. The algorithm performs better in datasets with fewer number of Jacobi edges.}
\label{table:performance}
\end{table*}

\subsection{Runtime performance}
The runtime performance of our algorithm depends on the number of topological features in the input dataset, specifically the number of Jacobi edges. \autoref{table:traversal} shows the computation time taken by individual traversal steps of the algorithm for different datasets and corresponding control polygons. $\#Cells$ represents the total number of tetrahedra in the domain, $\#Jacobi\ edges$ represents number of Jacobi edges, $\#Jacobi\ intersections$ represents number of Jacobi edges intersected by the line $L$ corresponding to a control polygon, $\#Tets\ containing\ FS$ represents number of tetrahedra that contain (intersect) the output fiber surface. Time taken by individual steps of the algorithm is specified in columns A, B, and C.

Ethanediol$^1$ and Ethanediol$^2$ are the same dataset represented by the same bivariate field ($f_1$: electron density, $f_2$ : reduced gradient ~\cite{Johnson2010JACS}), but associated with two different query control polygons. The time taken to compute the Jacobi intersections is similar for all control polygons of a particular dataset because it primarily depends on the number of Jacobi edges. We observe that this step is the computational bottleneck. The high total computation times for Tooth and Combustion is due to the large number of Jacobi edges in these datasets. The time taken for directed search  depends on the number of Jacobi edges that intersect the line $L$ because it corresponds to the number of directed searches that may be initiated. We observe that the location of the intersected Jacobi edges plays an equal if not more important role in determining the time for directed search. The blue control polygon in the Tooth dataset intersects with a larger number of Jacobi edges than the yellow polygon but requires lesser amount of time for directed search because several intersected Jacobi edges are visited during directed search initiated from another edge. As expected, we observe that the time taken for the restricted BFS step depends on the number of tetrahedra that contain the fiber surface. 

We collect these tetrahedra and pass them to TTK in order to reuse its implementation of the numerical computation of fiber surface within a tetrahedron. The time taken for extracting this collection of tetrahedra, storing them in a data structure, and transferring the collection to TTK is mentioned under the data transfer column in \autoref{table:traversal}. This data transfer step can be avoided if the proposed search algorithm is integrated in TTK, wherein the fiber surface can be computed as soon as a tetrahedron is marked during the restricted BFS step. \autoref{table:performance} shows the total time taken to compute the fiber surface \ie traversal time from \autoref{table:traversal} and fiber surface extraction time for the identified tetrahedra. TTK implementation traverses through all tetrahedra passed to it, computes the fiber surface section enclosed in a particular tetrahedron if the fiber surface passes through it. Since all tetrahedra identified by  our algorithm contain the fiber surface, TTK processes all tetrahedra.  The running time increases with the number of tetrahedra passed to TTK, refer \autoref{table:performance}. The running time of our algorithm is comparable with the exhaustive search in TTK, particularly when the number of Jacobi edges in the dataset is small (Ethanediol$^1$). The octree based domain search implemented in TTK is faster in all cases. For Ethanediol$^2$, the number of tetrahedra that contain the fiber surface is large and seem scattered throughout the octree, causing the octree version to perform worse than the exhaustive search. 

The restricted BFS step of our algorithm approach relies on triangulation data structure implemented in TTK. Accessing cell neighbors of each traversed cell increases the computation time by a large amount. The computation times are still comparable. For Thiophene-Quinoxaline, the two scalar fields are x:hole\_nto and y:particle\_nto. Hole\_nto represents the charge lost and particle\_nto represents the charge gained during  electronic transition of molecule. The green control polygons highlight the fiber surfaces corresponding to region that donated charge and red highlights the acceptor region in molecule~\cite{sharma2021segmentation}. Since the number of Jacobi edges are two orders of magnitude smaller than the number of cells, Jacobi set driven search performs fairly well. For Tooth ($f_1$: scalar field, $f_2$: gradient magnitude), the number of Jacobi edges and number of cells in the dataset are of same order. Computing Jacobi intersections  takes more time than traversing all cells in an exhaustive search.  The difference is lower for Combustion ($f_1$: scalar field, $f_2$: gradient magnitude) because the number of Jacobi edges are approximately one-tenth the total number of cells. The results are computed on a machine with 2.10GHz Intel(R) Xeon(R) Gold 6130, 32 core processor and 345 GB RAM. All runtimes are reported for a serial implementation. The runtimes are average of five runs out of seven. The runs with maximum and minimum runtime are discarded.

\begin{table*}[!t]
    \centering
\begin{tabular}{p{1.5cm}p{1cm}>{\raggedleft\arraybackslash}p{2cm}>{\raggedleft\arraybackslash}p{1.5cm}||>{\raggedleft\arraybackslash}p{1.25cm}>{\raggedleft\arraybackslash}p{1.25cm}>{\raggedleft\arraybackslash}p{1cm}||>{\raggedleft\arraybackslash}p{1.25cm}>{\raggedleft\arraybackslash}p{1.25cm}>{\raggedleft\arraybackslash}p{1cm}} 
&&&& \multicolumn{3}{c}{\small{FS component computation (msec)}} & \multicolumn{3}{c}{\small{FS computation (msec)}}\\
\hline
\small{Dataset} & \small{Control polygon} & \small{\#Tets containing FS component} & \small{\#Tets containing FS}  & \small{Jacobi intersections (A)} & \small{Directed Search (B)} & \small{Restricted BFS (C)} & \small{Jacobi intersections (A)} & \small{Directed Search (B)} & \small{Restricted BFS (C)}
\\
\hline
\multirow{4}{4em}{Ethanediol$^1$} & Blue & 2660 & 12804 & 30 & $0^*$ & 12 & 31 & $0^*$ & 57\\
& Green & 3524 & 3524 & 31 & $0^*$ & 16 & 31 & 23 & 15\\ 
& Red & 2372 & 4718 & 31 & $0^*$ & 10 & 31 & 11 & 21\\ 
& Black & 3796 & 7612 & 31 & $0^*$ & 17 & 30 & 5 & 33
\\  
\hline
\multirow{2}{4em}{Tooth} & Blue & 223076 & 223145 & 899 & $0^*$ & 1028 & 897 & 95 & 1021
\\ 
& Grey & 87185 & 87185 & 873 & $0^*$ & 399 & 885 & 351 & 397\\ \\  

\end{tabular}
\caption{Time taken by our algorithm to extract a component of the fiber surface. Time to compute the Jacobi intersections is independent of the component. Time required for directed search to locate the seed tetrahedron is negligible for component computation due to interactive selection of Jacobi edge. Time required to execute restricted BFS depends on the number of tetrahedra containing the fiber surface component, and is hence comparatively small relative to the total time. $0^*$ represents values smaller than 1 msec.}
\label{table:traversalFlexible}
\end{table*}

\begin{table*}[!t]
\begin{tabular}{p{1.5cm}p{1cm}>{\raggedleft\arraybackslash}p{2cm}>{\raggedleft\arraybackslash}p{1.5cm}||>{\raggedleft\arraybackslash}p{1.25cm}>{\raggedleft\arraybackslash}p{1.5cm}>{\raggedleft\arraybackslash}p{1cm}||>{\raggedleft\arraybackslash}p{1.25cm}>{\raggedleft\arraybackslash}p{1.5cm}>{\raggedleft\arraybackslash}p{1cm}} 

&&&&  \multicolumn{3}{c}{\small{FS component computation (msec)}} & \multicolumn{3}{c}{\small{FS computation (msec)}}\\
\hline
\small{Dataset} & \small{Control polygon} & \small{\#Tets containing FS component} & \small{\#Tets containing FS}  & \small{Traversal (T = A+B+C)} & \small{FS Extraction (D)} & \small{Total (T+D)} & \small{Traversal (T = A+B+C)} & \small{FS Extraction (D)} & \small{Total (T+D)}
\\
\hline
\multirow{4}{4em}{Ethanediol$^1$} & Blue & 2660 & 12804 & 42 & 4 & 46 & 88 & 21 & 109\\
& Green & 3524 & 3524 & 47 & 5 & 52 & 69 & 5 & 74\\ 
& Red & 2372 & 4718 & 41 & 4 & 45 & 63 & 8 & 71\\ 
& Black & 3796 & 7612 & 48 & 6 & 54 & 68 & 13 & 81
\\  
\hline
\multirow{2}{4em}{Tooth} & Blue & 223076 & 223145 & 1927 & 389 & 2316 & 2013 & 386 & 2399
\\ 
& Grey & 87185 & 87185 & 1272 & 151 & 1423 & 1633 & 151 & 1784\\ \\

\end{tabular}
\caption{Overall time to compute a particular fiber surface component is always lower than time taken to compute complete fiber surface due to interactive selection of Jacobi edges and due to the smaller number of tetrahedra that contain the selected fiber surface component.}
\label{table:performanceFlexible}
\end{table*}

\subsection{Applications}
The algorithm supports exploration of fiber surfaces in the vicinity of Jacobi edges and flexible computation of individual components of a fiber surface. The top row in \autoref{fig:flexibleEthane} shows the Jacobi edges projected onto the range space with four different control polygons. The Jacobi edges intersected by the line extended through the control polygon edge are highlighted with the same color as the control polygon edge. Visual inspection enables the user to identify edges that intersect the control polygon. Visualizing the intersecting Jacobi edges in the domain~(middle row) provides good insight into the spatial distribution of features in the data. This insight helps the user interactively select a Jacobi edge and explore its neighborhood using fiber surfaces. The insets show the selected control polygon edge in pink. The bottom row shows fiber surface components that are extracted by initiating a directed search from the selected Jacobi edges and launching a restricted BFS from the resulting seed cell. We observe that the selected components correspond to individual atoms and bond structure in Ethanediol. In \autoref{fig:flexibleTooth}, two control polygons are identified in a similar manner for the Tooth dataset resulting in two fiber surface components. The set of all intersecting Jacobi edges form a network in the vicinity of the output fiber surfaces. These networks or clusters act as a guide to select an intersecting Jacobi edge. Selecting one edge from each cluster is sufficient to explore the fiber surface component covered by other Jacobi edges within the connected cluster. The user may proceed with the exploration by visually analyzing these networks and selecting few Jacobi edges.

\autoref{table:traversalFlexible} presents a comparison between the time taken for flexible computation of a fiber surface component that corresponds to an interactively selected Jacobi edge against time taken to compute the complete fiber surface corresponding to a control polygon edge. The time taken for directed search is almost negligible in the former case because only one directed search is initiated for the interactively selected Jacobi edge. Time taken for restricted BFS is less in cases where the component is significantly smaller than the complete fiber surface. In \autoref{table:performanceFlexible}, fiber surface extraction time~(D) also depends on the number of tetrahedra that contain the fiber surface component or complete fiber surface. Total time~(T+D) for component calculation is less than time required for complete fiber surface computation thanks to a negligibly small directed search time and reduced time for steps~C and~D.

\section{Conclusions}
In this paper, we presented an output sensitive approach to compute the fiber surface for a  bivariate field defined over a tetrahedral mesh. The method does not require explicit computation of the Reeb space and runtime depends on the number of Jacobi edges. The runtime performance is good, with runtimes less than a few seconds, even though it is slower than TTK. A key benefit of the approach is that it enables exploration of fiber surfaces within the vicinity of Jacobi edges. It supports flexible computation of individual components of the fiber surface. In future work, we plan to improve the runtime of the algorithm via Jacobi set simplification and to devise methods for identifying interesting Jacobi edges, thereby avoiding the need for user interaction with the CSP or the projected range space. We observe that the Jacobi set intersection step dominates the runtime. We plan to explore space partitioning techniques like quad tree and employ parallelization strategies to reduce the time taken by this step. Further, we also plan to analyze the performance with respect to memory requirement and preprocessing time.

\acknowledgments{
This work is partially supported by MoE Govt. of India, a Swarnajayanti Fellowship from SERB India (DST/SJF/ETA-02/2015-16), an Indo-Swedish joint network project (DST/ INT/ SWD/VR/P-02/2019), and a Mindtree Chair research grant.}

\bibliographystyle{abbrv-doi}

\bibliography{references}
\end{document}